\documentstyle[12pt,preprint,prl,aps,amsfonts,floats]{revtex} 
\tighten 
\begin{document}  
%  
%\preprint{Imperial/TP/97-98/032, hep-th/9803157}
\title{The exact formula for neutrino oscillations }

\author{Massimo Blasone${}^{a,c}$, Peter A. Henning${}^b$  
and Giuseppe Vitiello${}^{c}$\thanks{e-mail: m.blasone@ic.ac.uk, 
P.Henning@gsi.de, vitiello@physics.unisa.it}}

\address{${}^{a}$ Blackett Laboratory, Imperial College, Prince Consort
Road, \\ London SW7 2BZ, U.K. } 
\address{${}^{b}$Institut f\"ur Kernphysik, TH Darmstadt,
         Schlo\ss gartenstra\ss e 9, \\ D-64289 Darmstadt, Germany}   
\address{${}^{c}$Dipartimento di Fisica dell'Universit\`a  
and INFN, Gruppo Collegato, Salerno \\  I-84100 Salerno, Italy} 
\maketitle 
\begin{abstract}  
We present the exact formula for neutrino oscillations. By resorting to
recent results of Quantum Field Theory of fermion mixing, we work out the 
Green's function formalism for mixed neutrinos.
The usual quantum
mechanical Pontecorvo formula is recovered in the relativistic limit. 
\end{abstract}  
\vspace{0.3cm}

P.A.C.S.: 14.60.Pq, 11.15.Tk, 12.15.Ff 

$$   $$

%%%%%%%%%%%%%%%%%%%%%%%%%%%%%%%%%%%%%%%%%%%%%%%%%%%%%%%%%%%%%%  
%\section{\bf Introduction}  
  
The mixing of mass eigenstates plays a relevant role in high  
energy physics: the family mixing problem of the 
standard model \cite{Fri}, neutrino oscillations \cite{BP78} 
and the solar neutrino problem \cite{Sol}, thermal field theory
\cite{h94rep} in hot hadronic matter  
are some examples where it must be considered. 
However, a thorough theoretical analysis of the mixing of fields in
Quantum Field Theory has not been carried out till 
ref.\cite{BV95}.

In this paper we present the exact formula for neutrino oscillations: this
is done by studying
the Green's  functions for mixed fermions. 
We show that in the study of neutrino oscillations it is necessary  
to take into account the full structure of the  
Heisenberg vacuum for mixed fermions, which exhibits a very rich
condensate-coherent state  
structure \cite{BV95,BHV96}.  
 
The result is an  
oscillation formula which differs from the usual one in the 
non-relativistic region, a result which was partially achieved in ref.  
\cite{BV95}.
With the present analysis of the full vacuum effects, we get, together  
with the energy dependent 
"squeezing" factor of the amplitude studied in
ref.\cite{BV95},  
also an additional term with a different oscillatory frequency: 
resonance is then possible also in vacuum for particular values  
of the masses and/or of the momentum, 
thus leading to a suppression or to  
an enhancement of the conversion probability. 

We consider the mixing problem for two Dirac fields   
with a bilinear interaction Lagrangian.   
We denote the fields with $\nu_e$ and $\nu_\mu$ (space-time dependence  
suppressed), referring thus explicitly to electron and muon neutrinos:   
however, the following results are valid for general Dirac fields.   
The Lagrangian we consider has the general form   
\begin{equation}\label{lag2}  
{\cal L} = {\bar \nu}_e\left(i\not\partial -m_{e}\right)\nu_e + {\bar   
  \nu}_\mu\left(i\not\partial - m_{\mu}\right)\nu_\mu  
- \; m_{e \mu} \;\left({\bar \nu}_{e}  
\nu_{\mu} + {\bar \nu}_{\mu} \nu_{e}\right)    
\;.\end{equation}  
Generalization to a higher number of flavors is straightforward.  
This Lagrangian can be fully diagonalized by  
substituting for the fields \cite{BP78}  
\begin{eqnarray} \nonumber  
\nu_{e}(x)   &=&\nu_{1}(x)\,\cos\theta +   
                        \nu_{2}(x)\,\sin\theta\\  
\label{rot1}  
\nu_{\mu}(x) &=&-\nu_{1}(x)\,\sin\theta   
                      + \nu_{2}(x)\,\cos\theta  
\;,\end{eqnarray}  
where $\theta$ is the mixing angle and  
$ m_{e} = m_{1}\cos^{2}\theta \,+\, m_{2} \sin^{2}\theta~$, 
$m_{\mu} = m_{1}\sin^{2}\theta \,+\, m_{2} \cos^{2}\theta~$,  
$m_{e\mu} =(m_{2}-m_{1})\sin\theta \cos\theta\,$. 
 $\nu_1$ and $\nu_2$ therefore  
are non-interacting, free fields, anticommuting with each  
other at any space-time point.  
They are explicitly given by  
\begin{equation}\label{2.2}\nu_{i}(x) =   
V^{-\frac{1}{2}} \sum_{{\bf k},r}
\left[u^{r}_{{\bf k},i}e^{-i\omega_{k,i} t} \alpha^{r}_{{\bf k},i}\:+   
v^{r}_{-{\bf k},i}e^{i\omega_{k,i} t}\beta^{r\dag }_{-{\bf k},i} 
\right] e^{i {\bf k}\cdot{\bf x}} , \; \; \; ~  i=1,2 \;.   
\end{equation}  
Here and in the following we use $t\equiv x_0$, when no misunderstanding 
arises.  
The vacuum for the $\alpha_i$ and $\beta_i$ operators is denoted by  
$|0\rangle_{1,2}$: 
$\; \; \alpha^{r}_{{\bf k},i}|0\rangle_{1,2}= \beta^{r }_{{\bf   
k},i}|0\rangle_{1,2}=0$.  
The anticommutation relations, completeness and
orthonormality relations are the usual ones (see ref.\cite{BV95}).
  
The fields $\nu_e$ and $\nu_\mu$ are thus   
completely determined through   
eq.(\ref{rot1}). In order to circumvent the difficulty of the  
construction of a Fock space for the mixed fields,  
which has been emphasized in   
a number of publications, e.g. \cite{Fuj,GKL92},  
it is useful to expand the flavor fields   
$\nu_e$ and $\nu_\mu$ in the same basis as $\nu_1$ and $\nu_2$,  
\begin{eqnarray}\label{exnue1}  
\nu_{e}^{\alpha}(x)    
&=& G^{-1}_{\theta}(t)\, \nu_{1}^{\alpha}(x)\, G_{\theta}(t)  
= V^{-\frac{1}{2}} \sum_{{\bf k},r}  
  \left[ u^{r,\alpha}_{{\bf k},1}e^{-i\omega_{k,1}t} \alpha^{r}_{{\bf
k},e}(t) +   
v^{r,\alpha}_{-{\bf k},1}
e^{i\omega_{k,1}t}\beta^{r\dag}_{-{\bf k},e}(t)  
\right]  e^{i {\bf k}\cdot{\bf x}} ,
\\ \label{exnum1} 
\nu_{\mu}^{\alpha}(x)   
&=& G^{-1}_{\theta}(t)\, \nu_{2}^{\alpha}(x)\, G_{\theta}(t)  
= V^{-\frac{1}{2}} \sum_{{\bf k},r}  
\left[ u^{r,\alpha}_{{\bf k},2}e^{-i\omega_{k,2}t} \alpha^{r}_{{\bf
k},\mu}(t) +   
v^{r,\alpha}_{-{\bf k},2}
e^{i\omega_{k,2}t}\beta^{r\dag}_{-{\bf k},\mu}(t)  
\right] e^{i {\bf k}\cdot{\bf x}} , 
\end{eqnarray}
where $G_{\theta}(t) = \exp\left[\theta \int d^{3}{\bf x}  
\left(\nu_{1}^{\dag}(x)\nu_{2}(x) -   
\nu_{2}^{\dag}(x) \nu_{1}(x)\right)\right]\,$   
is the generator of the mixing transformations (\ref{rot1}) 
\cite{BV95}.
Notice that the flavor annihilation operator, 
say $\alpha^{r}_{{\bf k},e}(t)= \int d^3{\bf x}\,u^{r \dag}_{{\bf k},1}  
e^{i \omega_{k,1}t - i {\bf k}\cdot{\bf x}}\nu_{e}(x)\,$,  
has contributions {\em also\/}  
from the anti-particle operator $\beta^{\dag}_2$ \cite{BV95}:  
\begin{equation}\label{exnue2} 
\alpha^{r}_{{\bf k},e}(t) \! = \! \cos\theta\,\alpha^{r}_{{\bf k},1}
\! + \! \sin\theta\;\sum_{s}\left(  
u^{r\dag}_{{\bf k},1} u^{s}_{{\bf k},2}  
e^{-i(\omega_{k,2}-\omega_{k,1})t}  
 \alpha^{s}_{{\bf k},2}\,+\,  
u^{r\dag}_{{\bf k},1} v^{s}_{-{\bf k},2}  
e^{i(\omega_{k,1}+\omega_{k,2})t} \beta^{s\dag}_{-{\bf k},2}\right)  \, .
\end{equation}
The term  containing 
$u^{r\dag}_{{\bf k},1} v^{s}_{-{\bf k},2}$, which is non-zero for
$m_1\neq m_2$, has been wrongly missed in
the usual treatment of mixing.
Relations similar to eq.(\ref{exnue2}) hold for the other   
operators \cite{BV95}. 
  
The bilinear mixed term of eq.(\ref{lag2}) generates   
four non-zero two point   
Green's functions for the mixed fields   
$\nu_e$, $\nu_\mu$. 
The crucial point is about how to compute these propagators: if one
(naively) uses the vacuum $|0\rangle_{1,2}$, one gets an inconsistent 
result (cf. eq.(\ref{pee2})). Let us show this by defining the
propagators as
\begin{equation}\label{matrix}  
\left(\begin{array}{cc}  
S_{ee}^{\alpha\beta}(x,y)&S_{\mu e}^{\alpha\beta}(x,y)  
\vspace{0.1cm} \\  
S_{e \mu}^{\alpha\beta}(x,y)&S_{\mu \mu}^{\alpha\beta}(x,y)  
\end{array} \right)\equiv   
\,_{1,2}\langle 0 |  \!\left(\begin{array}{cc}  
T \left[\nu^{\alpha}_{e}(x) \bar{\nu}^{\beta}_{e}(y)\right]&  
T \left[\nu^{\alpha}_{\mu}(x) \bar{\nu}^{\beta}_{e}(y)\right]  
\vspace{0.1cm} \\  
T \left[\nu^{\alpha}_{e}(x) \bar{\nu}^{\beta}_{\mu}(y)\right]&  
T \left[\nu^{\alpha}_{\mu}(x) \bar{\nu}^{\beta}_{\mu}(y)\right]  
\end{array} \right)\!|0 \rangle_{1,2}  \, ,
\end{equation}  
where $T$ denotes time ordering. Use of (\ref{rot1}) gives  
$S_{ee}$ in momentum representation as  
\begin{equation}\label{pert1}  
S_{ee}(k_0,{\bf k})=   
\cos^2\theta\; \frac{\not \! k + m_1}{k^2 - m_{1}^{2} + i\delta} \; +   
\;\sin^2\theta\; \frac{\not \! k + m_2}{k^2 - m_{2}^{2} + i\delta}  \; ,
\end{equation}  
which is just the weighted sum of the two propagators for the free   
fields $\nu_1$ and $\nu_2$. It coincides with the   
Feynman propagator obtained by resumming (to all orders)   
the perturbative series  
\begin{equation}\label{pert2} 
S_{ee} = S_e\left(1\,+\,m_{e\mu}^2 \,S_\mu S_e \,+\, m_{e\mu}^4 \,S_\mu   
S_e S_\mu S_e \,+\, ...\right) = 
S_e\left(1 - m_{e\mu}^2 \,S_\mu S_e \right)^{-1}\,,  
\end{equation} 
where the ``bare'' propagators are defined as $S_{e/\mu}=(\not \! k -   
m_{e/\mu} + i\delta)^{-1}$.    
In a similar way, one computes $S_{e\mu}$ and $S_{\mu e}$.

The transition amplitude for an electron neutrino created   
by $\alpha_{{\bf k},e}^{r \dag}$ at time $t=0$ 
going into the same  
particle at time $t$, is given by 
\begin{equation}\label{exnue4}  
{\cal P}^r_{ee}({\bf  k},t)= i   
u^{r \dag}_{{\bf k},1}e^{i\omega_{k,1}t}\,  
S^>_{ee}({\bf  k},t)\,\gamma^0 u^r_{{\bf k},1}\, ,  
\;.\end{equation}  
where the spinors $u_1$ and  $v_1$ form the basis in which the field
$\nu_e$ is expanded (cf. eq.(\ref{exnue1})).
Here, $S^>_{ee}$   
denotes the unordered Green's function (or Wightman function):
$i S^{>\alpha \beta}_{ee}(t,{\bf x};0,{\bf y}) =   
{}_{1,2}\langle0|\nu^{\alpha}_{e}(t,{\bf x}) \;  
\bar{\nu}^{\beta}_{e}(0,{\bf y})  |0\rangle_{1,2}$ .   
The explicit expression for $S^{>}_{ee}({\bf k},t)$ is  
\begin{equation}\label{gre2}  
S^{>\,\alpha\beta}_{ee}({\bf k},t) =  
 -i\sum_{r}   
 \left( \cos^2\!\theta\;  
 e^{-i\omega_{k,1} t}\; u^{r,\alpha}_{{\bf k},1}\;  
  \bar{u}^{r,\beta}_{{\bf k},1} \;   
 + \sin^2\!\theta\;  
 e^{-i\omega_{k,2} t} \;u^{r,\alpha}_{{\bf k},2} \;  
 \bar{u}^{r,\beta}_{{\bf k},2}\;  
 \right)   
\;.\end{equation}  
The amplitude eq.(\ref{exnue4}) 
is independent of the spin orientation and given by  
\begin{equation}\label{pee1}  
{\cal P}_{ee}({\bf k},t)  
  =\cos^2\!\theta\,  
 + \sin^2\!\theta\,|U_{\bf k}|^{2}\, e^{-i(\omega_{k,2}-\omega_{k,1}) t}  
\;.\end{equation}  

For different masses and for ${\bf k}\neq 0\,$, 
$|U_{\bf k}|$ is always $<1$ \cite{BV95}. Explicitly, 
\begin{equation}\label{uk}
|U_{\bf k}|^2 = \frac{1}{2} \sum\limits_{r,s}   
\left|u^{r\dag}_{{\bf k},2}\,u^{s}_{{\bf k},1}\right|^2  
=\left(\frac{\omega_{k,1}+m_{1}}{2\omega_{k,1}}\right)  
\!\left(\frac{\omega_{k,2}+m_{2}}{2\omega_{k,2}}\right)  
\!\left(1+\frac{|{\bf k}|^{2}}{(\omega_{k,1}+m_{1})  
(\omega_{k,2}+m_{2})}\right)^2 \,.
 \end{equation}
Notice that $|U_{\bf k}|^2\rightarrow 1$ 
in the relativistic limit $ |{\bf k}|\gg \sqrt{m_1 m_2}\;$: only in this
limit the squared modulus of ${\cal P}_{ee}({\bf k},t)$ 
does reproduce the Pontecorvo oscillation formula. 
  
Of course, it should be $\lim_{t\rightarrow 0^+} {\cal  
P}_{ee}(t)=1$. Instead, one obtains the unacceptable result 
\begin{equation}\label{pee2}  
{\cal P}_{ee}({\bf k},0^+)=  
\cos^2\!\theta + \sin^2\!\theta\, |U_{\bf k}|^2 < 1  
\;.\end{equation}  
This means that the {\em choice of the state\/} $|0\rangle_{1,2}$ in  
eq.(\ref{matrix}) and in the computation of the Wightman function  
is {\em not\/}  the correct one.  
The reason is that  $|0\rangle_{1,2}$ is not the vacuum state
for the flavor fields \cite{BV95}:
although the Lagrangian (\ref{lag2})   
is fully diagonalizable by means of the rotation (\ref{rot1}), this   
transformation does not leave invariant the vacuum $|0\rangle_{1,2}$.   
The mixing generator $G_\theta$ induces on it a $SU(2)$ coherent   
state structure, resulting in a new state,  
\begin{equation}\label{oemu}  
|0(t)\rangle_{e,\mu}\equiv G^{-1}_{\theta}(t)|0\rangle_{1,2}\,,  
\end{equation}  
which is the {\em flavor vacuum\/}, i.e. the vacuum state for the   
flavor operators $\alpha_{e/\mu}$, $\beta_{e/\mu}$ \cite{BV95}:  
$\alpha_{{\bf k},e/\mu}^{r}(t)|0(t)\rangle_{e,\mu}=\beta_{{\bf   
k},e/\mu}^{r}(t)|0(t)\rangle_{e,\mu}=0$. It is a   
condensate of four different fermion-antifermion pairs, of the form   
$\alpha_{{\bf k},i}^{r\dag}\beta_{-{\bf k},j}^{r\dag}$, with $i,j=1,2$ (see   
\cite{BV95} for the explicit form). Its   
non-perturbative nature results in the unitary inequivalence with   
the ``perturbative'' vacuum $|0\rangle_{1,2}$, in the infinite volume   
limit \cite{BV95}.  
An important point is represented by the time dependence of mixing   
generator, and consequently of the flavor vacuum. This is not surprising  
since the flavor states are not mass eigenstates and therefore the  
Poincar\'e structure of the flavor vacuum is lacking. 
  
We now show that the correct definition of the Green's function matrix
for the fields $\nu_e$, $\nu_\mu$ is the one which involves the   
flavor vacuum $|0\rangle_{e,\mu}$, i.e. 
\begin{equation}\label{matrix2}
\left(\begin{array}{cc}  
G_{ee}^{\alpha\beta}(x,y)&G_{\mu e}^{\alpha\beta}(x,y)  
\vspace{0.1cm}\\ 
G_{e \mu}^{\alpha\beta}(x,y)&G_{\mu \mu}^{\alpha\beta}(x,y)  
\end{array} \right)  
\equiv  \,_{e,\mu}\langle 0(y_0) | \! \left(\begin{array}{cc}  
T \left[\nu^{\alpha}_{e}(x) \bar{\nu}^{\beta}_{e}(y)\right]&  
T \left[\nu^{\alpha}_{\mu}(x) \bar{\nu}^{\beta}_{e}(y)\right]  
\vspace{0.1cm} \\ 
T \left[\nu^{\alpha}_{e}(x) \bar{\nu}^{\beta}_{\mu}(y)\right]&  
T \left[\nu^{\alpha}_{\mu}(x) \bar{\nu}^{\beta}_{\mu}(y)\right]  
\end{array} \right)\!|0(y_0) \rangle_{e,\mu}  \, .
\end{equation}
Notice that here the time argument $y_0$ (or, equally well, $x_0$)   
of the flavor ground state, is chosen to be equal on both sides   
of the expectation value. 
Indeed, we observe that transition matrix elements of the type  
${}_{e,\mu}\langle0|\alpha_e\,\exp\left[ -i H t\right]\,  
\alpha^{\dag}_e | 0\rangle_{e,\mu}$, where 
$H$ is the Hamiltonian, do not represent    
physical transition amplitudes: they actually vanish (in the infinite
volume limit) due to the unitary inequivalence of flavor vacua at
different times (see below). 
Therefore the comparison of states at different times  
necessitates a {\em parallel transport\/} of these states to  
a common point of reference. The definition (\ref{matrix2}) includes  
this concept of parallel transport, which is a sort of
``gauge fixing'': a rich geometric structure underlying the 
mixing transformations (\ref{rot1}) is thus uncovered. In a forthcoming
publication we will discuss these geometrical aspects of field mixing,
including Berry phase and gauge structure. 
  
In the case of $\nu_e \rightarrow \nu_e$   
propagation, we now have (for ${\bf k}=(0,0,|{\bf k}|)$):  
\begin{eqnarray} \nonumber  
G_{ee}(k_0,{\bf k})&=&S_{ee}(k_0,{\bf k})\; + \;  2\pi\,i\, 
\sin^2\theta \Big[|V_{{\bf k}}|^2 
\; (\not \! k + m_2)\;\delta(k^2 - m_2^2)     
\\  \label{diff}  
&&  \;-\; |U_{{\bf k}}| |V_{{\bf k}}| 
\;\sum_{r} \left(\epsilon^r u^{r}_{{\bf k}   
,2}\;\bar{v}^{r}_{-{\bf k},2} \; \delta(k_0 - \omega_2 )\; + \;\epsilon^r   
v^{r}_{-{\bf k},2} \;\bar{u}^{r}_{{\bf k} ,2}\;\delta(k_0 + \omega_2 )   
\right)\Big]  \, ,
\end{eqnarray}  
where we used $\epsilon^r=(-1)^r\,$ and 
$|V_{\bf k}|=\sqrt{1-|U_{\bf k}|^2}$. Comparison of eq.(\ref{diff}) 
with eq.(\ref{pert1}) 
shows that the difference between the   
full and the perturbative propagators is in the imaginary part.  
  
The Wightman functions for an electron neutrino are   
$i G^{>\alpha \beta}_{ee}(t,{\bf x};0,{\bf y}) =   
{}_{e,\mu}\langle0|\nu^{\alpha}_{e}(t,{\bf x}) \;  
\bar{\nu}^{\beta}_{e}(0,{\bf y})  |0\rangle_{e,\mu}$, 
and $i G^{>\alpha \beta}_{\mu e} (t,{\bf x};0,{\bf y}) =   
{}_{e,\mu}\langle0|   
\nu^{\alpha}_{\mu}(t,{\bf x}) \; \bar{\nu}^{\beta}_{e}(0,{\bf y})  
|0\rangle_{e,\mu}$. 
These are conveniently expressed in terms of 
anticommutators at different times as  
\begin{eqnarray}\label{gfu2} 
i G^{>\alpha\beta}_{ee}({\bf  k},t) &=& \sum_{r}  
\left[u^{r,\alpha}_{{\bf k},1}\,   
\bar{u}^{r,\beta }_{{\bf k},1}  
\left\{\alpha^r_{{\bf k},e}(t),\alpha^{r\dag}_{{\bf k},e}\right\} 
e^{-i\omega_{k,1}t} +  \,
v^{r,\alpha}_{-{\bf k},1}\,   
\bar{u}^{r,\beta}_{{\bf k},1}  
\left\{\beta^{r\dag}_{-{\bf k},e}(t),\alpha^{r\dag}_{{\bf k},e}
\right\}e^{i\omega_{k,1}t} \right],
\\ \label{gfu3}  
i G^{>\alpha\beta}_{\mu e}({\bf  k},t) &=& \sum_{r}  
\left[u^{r,\alpha}_{{\bf k},2}\, 
\bar{u}^{r,\beta }_{{\bf k},1}  
\,\left\{\alpha^r_{{\bf k},\mu}(t),\alpha^{r\dag}_{{\bf k},e}\right\}
e^{-i\omega_{k,2}t}  + 
\,v^{r,\alpha}_{-{\bf k},2}\, \bar{u}^{r,\beta}_{{\bf k},1}  
\,\left\{\beta^{r\dag}_{-{\bf k},\mu}(t),  
\alpha^{r\dag}_{{\bf k},e}\right\}e^{i\omega_{k,2}t} \right] .
\end{eqnarray}
Here and in the following 
$\alpha^{r\dag}_{{\bf k},e}$ stands for $\alpha^{r\dag}_{{\bf k},e}(0)$.
We now have four distinct transition amplitudes, given by anticommutators
of flavor operators at different times:
\begin{eqnarray} \nonumber  
{\cal P}^r_{ee} ({\bf  k},t)  
 &\equiv& i \,u^{r \dag}_{{\bf k},1}{ e}^{i\omega_{k,1}t}\,  
 G^>_{ee}({\bf  k},t)\,\gamma^0 u^r_{{\bf k},1} =   
\left\{\alpha^r_{{\bf k},e}(t),\alpha^{r\dag}_{{\bf k},e} \right\}  
\\ \label{pee3}  
 &=&  \cos^{2}\!\theta\,  
 + \sin^2\!\theta\,\left[ |U_{\bf k}|^{2} { e}^{-i  
 (\omega_{k,2}-\omega_{k,1}) t}  
     + |V_{\bf k}|^{2} e^{i(\omega_{k,2}+\omega_{k,1}) t}\right],  
\\ \nonumber 
\\ \nonumber  
{\cal P}^r_{\bar{e}e} ({\bf  k},t)  
 &\equiv&i \,v^{r \dag}_{-{\bf k},1}{ e}^{-i\omega_{k,1}t}\,  
 G^>_{ee}({\bf  k},t)\,\gamma^0 u^r_{{\bf k},1} =  
  \left\{\beta^{r\dag}_{-{\bf k},e}(t),\alpha^{r\dag}_{{\bf k},e} \right\}  
\\ \label{pee4}  
 &=&  \epsilon^r\,|U_{\bf k}| |V_{\bf k}|\,\sin^{2}\!\theta\,  
 \left[  e^{i (\omega_{k,2}-\omega_{k,1})t}\;  
-\; e^{-i (\omega_{k,2}+\omega_{k,1})t} \right] , 
\\ \nonumber 
\\ \nonumber  
{\cal P}^r_{\mu e}({\bf  k},t)&\equiv& i \,u^{r   
\dag}_{{\bf k},2}e^{i\omega_{k,2}t}\,  
 G^>_{\mu e}({\bf  k},t)\,\gamma^0 u^r_{{\bf k},1} =   
\left\{\alpha^r_{{\bf k},\mu}(t),\alpha^{r\dag}_{{\bf k},e} \right\}   
\\ \label{pee5}  
&=&\;|U_{\bf k}|\;\cos\!\theta\;\sin\!\theta \left[1\;-\;  
 e^{i (\omega_{k,2}-\omega_{k,1}) t}\right],  
\\ \nonumber 
\\ \nonumber  
{\cal P}^r_{\bar{\mu} e}({\bf  k},t)&\equiv&  
i \,v^{r \dag}_{-{\bf k},2}e^{-i\omega_{k,2}t}\,  
 G^>_{\mu e}({\bf  k},t)\,\gamma^0 u^r_{{\bf k},1} =   
\left\{\beta^{r\dag}_{-{\bf k},\mu}(t),\alpha^{r\dag}_{{\bf k},e} \right\}   
\\ \label{pee6}  
&=&\;\epsilon^r\;|V_{\bf k}|\;\cos\!\theta\;\sin\!\theta   
\left[1\; -\; e^{-i (\omega_{k,2}+\omega_{k,1})t} \right]  
\;.\end{eqnarray}  
All other anticommutators with $\alpha^{\dag}_e $  
vanish. 
The probability amplitude is now  
correctly normalized:  $lim_{t\rightarrow 0^+}  
{\cal P}_{ee}({\bf  k},t)=1$, and ${\cal P}_{\bar{e}e}$, 
${\cal P}_{\mu e}$, ${\cal P}_{\bar{\mu} e}$ go to zero in the same 
limit $t\rightarrow 0^+\,$. Moreover, 
\begin{equation}\label{cons}  
\left|{\cal P}^r_{ee}({\bf  k},t)\right|^2 +   
\left|{\cal P}^r_{\bar{e}e}({\bf  k},t)\right|^2 +  
\left|{\cal P}^r_{\mu e}({\bf  k},t)\right|^2 +  
\left|{\cal P}^r_{\bar{\mu}e}({\bf  k},t)\right|^2 =1  
\,,\end{equation}  
as the conservation of the total probability requires. We also note  
that the above transition probabilities are independent of the spin 
orientation.  
 
For notational simplicity, we now   
drop the momentum and spin indices. The momentum   
is taken to be aligned along the quantization axis,
${\bf k}=(0,0,|{\bf k}|)$. It is also understood that antiparticles   
carry opposite momentum to that of the particles. 
At time $t=0$, the vacuum state is $|0\rangle_{e,\mu}$  
and the one electron neutrino state is   
\begin{equation}\label{h1}  
|\nu_e \rangle \equiv \alpha_{e}^{\dag}|0\rangle_{e,\mu} = \left[   
\cos\theta\,\alpha_{1}^{\dag} +  
|U|\; \sin\theta\;\alpha_{2}^{\dag} - \epsilon \; |V| \,\sin\theta \;  
\alpha_{1}^{\dag}\alpha_{2}^{\dag}\beta_{1}^{\dag}  
 \right] |0\rangle_{1,2} \,.  
\end{equation}  
In this state a multiparticle component is present,   
disappearing in the relativistic limit $|{\bf k}|\gg \sqrt{m_1m_2}\,$: in this   
limit the (quantum-mechanical) Pontecorvo state is recovered.  
The time evoluted of $|\nu_e\rangle$ is given by $|\nu_e (t)\rangle
= e^{-iH t} |\nu_e\rangle$.  

Notice however that the flavor vacuum   
$|0\rangle_{e,\mu}$ is not eigenstate of the free Hamiltonian $H$. 
It ``rotates'' under the action of the time  
evolution generator: one indeed finds 
$\lim_{V \rightarrow \infty}\;_{e,\mu}\langle 0\;|\;0(t)\rangle_{e,\mu} =  
0$. Thus at different times we have unitarily inequivalent flavor  
vacua (in the limit $V\rightarrow \infty$): this expresses  
the different particle content of these (coherent) states and it is  
direct consequence of the fact that flavor states are not mass  
eigenstates. 

As already observed, this implies that we cannot directly compare flavor
states at different times. However we can consider the flavor charge
operators, defined as  
$Q_{e/\mu}\equiv \alpha_{e/\mu}^{\dag}\alpha_{e/\mu} -  
\beta^{\dag}_{e/\mu}\beta_{e/\mu}$.  We then have 
\begin{eqnarray} \label{charge1} 
&&\;_{e,\mu}\langle 0(t)|Q_e|  
0(t)\rangle_{e,\mu}\; =\;_{e,\mu}\langle 0(t)|Q_{\mu}|  
0(t)\rangle_{e,\mu} \; =\; 0 \, ,
\\ \label{charge2} 
&& \langle \nu_e(t)|Q_e| \nu_e(t)\rangle\; = \; 
\left|\left\{\alpha_{e}(t), \alpha^{\dag}_{e} \right\}\right|^{2} \;+ 
\;\left|\left\{\beta_{e}^{\dag}(t),\alpha^{\dag}_{e} \right\}\right|^{2}
\, , 
\\ \label{charge3} 
&&\langle \nu_e(t)|Q_{\mu}| \nu_e(t)\rangle\; = \; 
\left|\left\{\alpha_{\mu}(t), \alpha^{\dag}_{e} \right\}\right|^{2} \;+ 
\;\left|\left\{\beta_{\mu}^{\dag}(t),\alpha^{\dag}_{e}
\right\}\right|^{2}\, .
\end{eqnarray}  

Charge conservation is ensured at any time: 
$\langle \nu_e(t)|\left(Q_e\;+\;Q_{\mu}\right)| \nu_e(t)\rangle\; = \; 1$
and the oscillation  
formula readily follows as
\begin{eqnarray} \label{enumber}  
P_{\nu_e\rightarrow\nu_e}({\bf k},t)&=& \left|\left 
\{\alpha^{r}_{{\bf k},e}(t),  
\alpha^{r \dag}_{{\bf k},e} \right\}\right|^{2} \;+  
\;\left|\left\{\beta_{{-\bf k},e}^{r \dag}(t), 
\alpha^{r \dag}_{{\bf k},e} \right\}\right|^{2}  
\\ \nonumber 
&=& 1 - \sin^{2}( 2 \theta) \left[ |U_{{\bf k}}|^{2} \;   
\sin^{2} \left( \frac{\omega_{k,2} - \omega_{k,1}}{2} t \right)  
+|V_{{\bf k}}|^{2} \;  
\sin^{2} \left( \frac{\omega_{k,2} + \omega_{k,1}}{2} t \right) \right]
\, , 
\\ \nonumber {} 
\\ \label{munumber}  
P_{\nu_e\rightarrow\nu_\mu}({\bf k},t)&=&  
\left|\left\{\alpha^{r}_{{\bf k},\mu}(t), 
\alpha^{r \dag}_{{\bf k},e} \right\}\right|^{2} \;+  
\;\left|\left\{\beta_{{-\bf k},\mu}^{r \dag}(t), 
\alpha^{r \dag}_{{\bf k},e} \right\}  
\right|^{2}  
\\ \nonumber 
&=&  \sin^{2}( 2 \theta)\left[ |U_{{\bf k}}|^{2} 
\;  \sin^{2} \left( \frac{\omega_{k,2} - \omega_{k,1}}{2} t \right)   
+|V_{{\bf k}}|^{2} \;  
\sin^{2} \left( \frac{\omega_{k,2} + \omega_{k,1}}{2} t \right) \right]
\, .
\end{eqnarray}  

This result is exact.
There are two differences with respect to the usual formula for neutrino
oscillations: the amplitudes are energy dependent, and there is an
additional oscillating term.
For $|{\bf k}|\gg\sqrt{m_1m_2}$, 
$|U_{{\bf k}}|^{2}\rightarrow 1$ 
and  $|V_{{\bf k}}|^{2}\rightarrow 0$ 
and the traditional oscillation formula is
recovered. 
However, also in this case we have that the neutrino state remains a
coherent state, thus phenomenological implications of our analysis 
are possible also for relativistic neutrinos. 
Further work in this direction is in progress. 
\smallskip
%%%%%%%%%%%%%%%%%%%%%%%%%%%%%%%%%%%%%%%%%%%%%%%%%%%%%%%%%%%%%%%%%%%%%%%%%%%

This work is partially supported by MURST, INFN, and by a network
supported by the European Science Foundation.
 
%%%%%%%%%%%%%%%%%%%%%%%%%%%%%%%%%%%%%%%%%%%%%%%%%%%%%%%%%%%%%%%%%%%%%%%%%%%%  

\end{document}